# New Diluted ferromagnetic semiconductor Li(Zn,Mn)P with decoupled charge and spin doping

Z. Deng[1], K. Zhao[1], B.Gu[2], W.Han[1], J.L. Zhu[1], X. C. Wang[1], X. Li[1], Q.Q. Liu[1], R.C. Yu[1], T. Goko[3], B. Frandsen[3], L. Liu[3], Jinsong Zhang[4], Yayu Wang[4], F. L. Ning[5], S. Maekawa[2], Y.J. Uemura[3], C.Q.Jin[1*]

[1]*Beijing National Laboratory for Condensed Matter Physics, and Institute of Physics, Chinese Academy of Sciences, Beijing 100190, China*
[2]*Advanced Science Research Center, Japan Atomic Energy Agency, Tokai 319-1195；CREST, Japan Science and Technology Agency, Sanbancho, Tokyo 102-0075, Japan*
[3]*Department of Physics, Columbia University, New York, New York 10027, USA*
[4]*Department of Physics, Tsinghua University, Beijing 100084, China*
[5]*Department of Physics, Zhejiang University, Hangzhou 310027, China*

We report the discovery of a new diluted magnetic semiconductor, Li(Zn,Mn)P, in which charge and spin are introduced independently via lithium off-stoichiometry and the isovalent substitution of $Mn^{2+}$ for $Zn^{2+}$, respectively. Isostructural to (Ga,Mn)As, Li(Zn,Mn)P was found to be a p-type ferromagnetic semiconductor with excess Lithium providing charge doping. First principles calculations indicate that excess Li is favored to partially occupy the Zn site, leading to hole doping. Ferromagnetism is mediated in semiconducting samples of relative low mobile carriers with a small coercive force, indicating an easy spin flip.

## INTRODUCTION

Spintronic devices utilize the electron's charge and spin degrees of freedom to achieve novel quantum functionalities. Diluted magnetic semiconductors (DMS) constitute an important category of spintronic materials that have the potential to be successfully incorporated into the existing semiconductor industry. The prototypical DMS (Ga,Mn)As,[1] discovered in the 1990s, accomplishes spin and charge doping simultaneously through the heterovalent substitution of the magnetic ion $Mn^{2+}$ for $Ga^{3+}$. Two challenges have presented themselves in this material. First, the heterovalent nature of this "integrated spin/charge" doping results in severely limited chemical solubility in (Ga,Mn)As, restricting specimen fabrication to metastable thin films by molecular beam epitaxy; second, the simultaneous spin and charge doping precludes the possibility of individually tuning the spin and charge degrees of freedom.

A new type of ferromagnetic DMS that overcomes both of these challenges was recently discovered.[2] Li(Zn,Mn)As utilizes excess Li concentration to introduce hole carriers, while independently making the isovalent substitution of $Mn^{2+}$ for $Zn^{2+}$ to achieve local spin doping. With no heterovalent substitutions to restrict chemical solubility, bulk specimens of Li(Zn,Mn)As were successfully fabricated.[2] Shortly thereafter, another new bulk ferromagnetic DMS with independent charge and spin doping, $(Ba,K)(Zn,Mn)_2As_2$, was found with Curie temperature ($T_C$) up to 180K. Holes are doped via ($Ba^{2+}$, $K^{1+}$) substitution, while spins are introduced by isovalent ($Zn^{2+}$, $Mn^{2+}$) substitution.[3] These new DMS materials have the added advantage of being structurally very similar to the recently discovered iron-pnictide high-temperature superconductors, opening up the possibility of creating novel junctions between superconducting, semiconducting, and magnetic materials.

It is noted that previously DMS are successfully obtained on II-VI based compounds such as (Zn,Mn)Te, (Cd,Mn)Te,[4,5] in which the valence of group II cations is identical to that of doped magnetic element Mn, resulting in precise DMS properties. For the Li(Zn,Mn)As DMS system, a separate carrier doping mechanism was introduced by Lithium stoichiometry, in additional to spin injection via isovalent $Mn^{2+}$ substitution to $Zn^{2+}$. This provides more freedom to further precisely tune DMS properties. However one drawback of Li(Zn,Mn)As is its use of the toxic element As. In the present work, we report the synthesis of a new DMS in which As has been replaced by nontoxic P as shown in Fig. 1(a). We found that Li(Zn,Mn)P exhibits soft ferromagnetic behavior with a relative lower carrier density than those for Li(Zn,Mn)As and (Ga,Mn)As),[2,6–8] but with comparable $T_C$, offering the advantage of much improved semiconductive behavior and the potential for higher $T_C$ via increased carrier concentration.

**EXPERIMENTAL**

Polycrystalline specimens were prepared as described in Ref. 2 and 3. High purity starting materials with molar amounts proportional to the nominal element concentrations were pressed into a pellet and loaded into a Ta tube with high purity argon. The Ta tube, which prevented the evaporation of Li, was then sealed into an evacuated quartz tube. All procedures were performed in a glove box under high purity argon atmosphere. The samples were heated to 900°C and held for several days before the temperature was slowly decreased. The specimens were characterized by X-ray powder diffraction on a PaNalytical X'pert diffractometer operating with Cu-$K_\alpha$ radiation. Lattice parameters were determined via Rietveld analysis using the GSAS software package. The real atomic rations of heavy elements such as Zn, Mn or As measured from energy dispersive analysis (EDX) are nearly the same as the nominal ones. Such as the Mn:(Zn+Mn) is about 0.032, 0.051, 0.074 for nominal $Li1_{1.02}(Zn_{0.97}Mn_{0.03})P$, $Li1_{1.02}(Zn_{0.95}Mn_{0.05})P$ and $Li1_{1.02}(Zn_{0.93}Mn_{0.07})P$, respectively. It is rather difficult to measure Li concentration since it is a very light element. But its concentration evolution can be inferred from lattice parameter change. Successful Li and Mn doping are supported by continuing lattice constant increasing as shown in Fig. 1(b). The concentrations of Li and Mn in the paper therefore use the nominal composition. DC magnetization measurements were performed on a Quantum Design Superconducting Quantum Interference Device-Vibrating Sample Magnetometer (SQUID-VSM). Resistivity and Hall effect measurements were carried out on a Quantum Design Physical Property Measurement System (PPMS) using the four-probe method and the Hall bar method, respectively.

## RESULTS AND DICUSSION

### I. Experimental results

Step-scanning powder X-ray diffraction measurements showed that pristine LiZnP crystallizes into a structure similar to zinc-blend type GaAs with space group F-43m,[2] as shown in Figure 1(a). The refined lattice parameter of LiZnP is $a = 5.7564$Å.[9,10] We found that chemically stable bulk crystals of $Li_{1+y}(Zn_{1-x}Mn_x)P$ can be obtained for excess Li with $y \leq 0.15$ and Mn concentrations $x$ up to at least 0.1. As shown in Figure 1(b), the lattice parameter evolved systematically with Li and Mn concentrations, suggesting successful chemical doping.

The temperature-dependent magnetization *M(T)* and field-dependent magnetization *M(H)* were measured for $Li_{1+y}(Zn_{1-x}Mn_x)P$ with $-0.05 \leq y \leq 0.07$ and $x = 0, 0.03, 0.06, 0.1$. The specimens with deficient Li showed spurious weak ferromagnetic signals above room temperature caused by the ferromagnetic impurity MnP.[11] Specimens with excess Li, on the other hand, demonstrated robust ferromagnetism, as displayed in Fig. 2. Figure 2(a) shows *M(T)* for $Li_{1.04}(Zn_{1-x}Mn_x)P$ specimens, $x = 0, 0.03, 0.06, 0.1$, in an applied field of 100Oe in both zero-field cooling (ZFC) and field cooling (FC) modes. $T_C$, determined by the projected line method, clearly increases up to 34K, as shown in Figure 2(b). Above $T_C$, the susceptibility $\chi$ can be fit to a Curie-Weiss law (Fig. 2(a), 2(b)), $(\chi-\chi_0)^{-1} = (T-\theta)/C$, where $\chi_0$ is a temperature-independent term, $C$ is the Curie constant, and $\theta$ is the Weiss temperature. The positive value of $\theta$ found for $Li_{1.04}(Zn_{1-x}Mn_x)P$ (Fig. 2(b)) indicates a ferromagnetic interaction between $Mn^{2+}$ ions.

The effective paramagnetic moment ($M_{eff}$) obtained from the Curie constant (solid purple circles in Figure 2(b)) decreases with increasing Mn doping, a trend also found in other systems doped with magnetic ions.[12–14] Extrapolation of $M_{eff}$ to lower Mn concentrations yields a value of approximately 5.9 $\mu_B$/Mn, as expected for the fully high-spin oriented $Mn^{2+}$ ion. Figure 2(b) also shows the saturation moment ($M_{sat}$) per Mn in an applied field of 500Oe, found to be about 1~2 $\mu_B$/Mn. For ferromagnetic (Ga,Mn)As, Li(Zn,Mn)As and $(Ba,K)(Zn,Mn)_2As_2$, $M_{sat}$ are about 2~4 $\mu_B$/Mn,[15] 1~3$\mu_B$/Mn,[2] 1~2$\mu_B$/Mn,[3] respectively, which are comparable to that of Li(Zn,Mn)P. As with $M_{eff}$, $M_{sat}$ decreases with increasing Mn concentration, likely due to competition between antiferromagnetic coupling of nearest neighbor Mn moments and ferromagnetic coupling of Mn moments mediated by the doped hole carriers. Figure 2(c) shows the hysteresis curves for $Li_{1.04}Zn_{0.9}Mn_{0.1}P$, indicating soft ferromagnetic behavior with a very small coercive field ($H_C$) of ~50 Oe. For potential spintronics application, spin will stand for memory unit. The lower the magnetic coercive force is, the easier spin can be flipped. This small coercive force of Li(ZnMn)P DMS could be advantageous for spin flip manipulation, benefiting memory processing for potential applications.

LiZnP was reported to be a semiconductor with a direct band gap around 2.04 eV.[9] Figure 3(a) shows the resistivity and carrier concentration of $Li_{1.04}Zn_{0.9}Mn_{0.1}P$ from 5K - 300K. The resistivity value was diverging and too large at low temperature. Any small misalignment of the two Hall contacts would pick up a longitudinal resistivity signal and this brought large trouble in the Hall effect measurement. As shown in the inset of Fig. 3(a), even 100K Hall curve became difficult for measurement. Therefore no anomalous Hall effect signal was detected so far because of technical problem. However Hall effect measurement of relative high temperature already told us the carrier type and concentration. Nevertheless, the hysteretic behavior in the magnetoresistance (MR) curve below Cuire temperature as shown in Fig.3 provides evidence on for the coupling between carriers and local moments. The conclusion that ferromagnetic order was mediated by very low carrier concentration is what we can present in this paper. Present paper just reports first step of new DMS material. Higher quality single crystal or epitaxil film which we are working on will solve the problem.

As plotted in Fig. 3(a), the resistivity obviously increases with decreasing temperature, whereas the mobile hole concentration decreases. This is indicative of typical semiconducting behavior. Somewhat surprisingly, $Li_{1+y}(Zn,Mn)P$, $0<y<0.7$, is found to exhibit p-type behavior, not the n-type behavior that one would expect assuming that excess Li provides additional electrons. This is explained by first principles calculations, which indicate that the excess $Li^{1+}$ ions are thermodynamically favored to occupy the $Zn^{2+}$ sites (see the following section), thereby rendering Li(Zn,Mn)P a p-type DMS.

Previous studies of $(Ga_{1-x}Mn_x)As$ demonstrated that for $0.012 \leq x \leq 0.03$, ferromagnetic order can be achieved in the full volume fraction while the system still shows semiconducting transport behavior,[15] implying that the paramagnetic to ferromagnetic quantum transition occurs at a different charge/spin concentration than the semiconductor to metal quantum transition. The present work confirms the same situation for Li(Zn,Mn)P. The hole concentration calculated from Hall effect measurements of charge/spin co-doped $Li_{1.04}Zn_{0.9}Mn_{0.1}P$ (inset of Fig. 3(a)) is larger than that of pristine LiZnP by more than a factor of 10, but still results in semiconducting transport behavior.[9] Nonetheless, this small hole concentration is evidently sufficient to mediate ferromagnetic coupling of the dilute Mn moments and induce magnetic order. It is likely that the doped holes are not fully delocalized but have sufficient spread in real space to mediate magnetic coupling between neighbouring Mn spins, thereby facilitating the buildup of a percolating magnetic network.

Figure 3(b) compares the hole concentrations and ferromagnetic transition temperatures of Li(Zn,Mn)P to those of other DMS systems. The hole concentration of $Li_{1.04}Zn_{0.9}Mn_{0.1}P$ is more than two orders of magnitude smaller than that of typical metallic DMS ferromagnets.[2,3,14,16,17] The relationship between hole concentration and $T_C$ exhibited by the other systems suggests that further charge and spin doping would cause Li(Zn,Mn)P to become metallic and magnetically order at a higher $T_C$.[18–22] We are currently working on optimizing the materials processing to show this behavior.

Magnetotransport measurements performed on $Li_{1.04}Zn_{0.9}Mn_{0.1}P$ are shown in Figure 3(c). Several types of effects can produce a magnetoresistance in magnetic semiconductors. For a DMS, the negative magnetoresistance usually results from the reduction of spin-dependent scattering by aligning the spins in the applied field.[1,14] However, as shown in Figure 2(c) and Figure 3(c), the negative magnetoresistance is far from saturation in rather high magnetic field, in which spin orientation is fully aligned. In this condition, weak localization effects presumably contribute to negative magnetoresistance.[23] No signature of a metal-insulator phase transition could be found at the Curie point.[15,16,24] The resistivity $\rho$ increases monotonically with decreasing temperature, showing a rapid rise below $T_C$. Hysteresis is observed in $\rho(H)$ for low fields (~50Oe) and low temperatures, corresponding closely to the behavior of the magnetization $M(H)$ (see Fig. 2(c)).

**II. Theoretical analysis**

We studied the electronic state in Li(Zn,Mn)P with excess Li. We found that (i) for the compound the excess Li atoms prefer to occupy Zn-substitutional sites $Li_{Zn}$, and thus create the p-type carriers; (ii) the ferromagnetic correlations between Mn ions develop for the case of p-type carriers, and the corresponding effective exchange constant is smaller in Li(Zn,Mn)P than that in Li(Zn,Mn)As leads to lower Currie Temperature.

To study the stable state for excess Li atoms in Li(Zn,Mn)P compound, we calculated the electronic structures by using the density functional theory (DFT) implemented in the code QUANTUM ESPRESSO.[25] The exchange-correlation interactions are described by the Perdew-Burke-Ernzerhof generalized gradients approximation (GGA), and the electron-ion interactions are represented by the Vanderbilt ultrasoft pseudopotentials. We calculate the formation energy for the interstitial site $Li_I$ and the Zn substitutional site $Li_{Zn}$, respectively. Since Mn at Zn-substitutional sites $Mn_{Zn}$ do not introduce any carriers, we study the excess Li in LiZnP. Excess Li atom at interstitial sites $Li_I$ contributes one electron while $Li_{Zn}$ contributes one hole to the system. Our DFT calculations show that $Li_{Zn}$ makes the formation energy to be lower than $Li_I$ case as shown in Table 1. It means the carriers will be holes.

To study the magnetic correlations between Mn in Li(Zn,Mn)P, we take the two-step calculations by our combined QFT+QMC (quantum Monte Carlo) method.[26,27] First, the one-particle parts of the Anderson impurity model are formulated within the DFT for determining the host band structure and the impurity-host hybridization. Second, the correlation parts of the Anderson impurity model at finite temperatures are calculated by the QMC method. The calculation details will be published elsewhere.[28] In p-type Li(Zn,Mn)P, ferromagnetic correlations between Mn impurities are obtained by our QMC calculats. In addition, we find that the effective exchange coupling in Li(Zn,Mn)P is smaller than that in Li(Zn,Mn)As.[29] By simple molecular-field theory, larger $T_C$ is expected for larger exchange coupling. This maybe the reason why a lower Curie temperature is obtained in Li(Zn,Mn)P than that in Li(Zn,Mn)As.

**IV CONCLUSIONS**

In summary, a new bulk diluted magnetic semiconductor Li(Zn,Mn)P was successfully synthesized with decoupled spin and charge doping. Li(Zn,Mn)P is a soft magnet with a relatively small coercive field. Ferromagnetic order sets in with lower hole concentrations than typical metallic DMS ferromagnets, indicating the potential to further enhance of $T_C$ with more itinerant carriers. This new DMS will contribute to the further development of new semiconductor materials and spintronic devices based on the individual tuning of spin and charge.

**ACKNOWLEDGEMENTS**

This work was funded by the Chinese NSF and Ministry of Science and Technology (MOST) through research projects; the US NSF PIRE (Partnership for International Research and Education: OISE-0968226) and DMR- 1105961 projects at Columbia; the JAEA Reimei project at IOP, Columbia, PSI, McMaster and TU Munich; and NSERC and CIFAR at McMaster.

TAB. 1. Formation energy for excess Li atom at different sites, obtained by DFT calculations. By Ref. 25, the formation energy is given by Eformation = ET - $n_{Li}\mu_{Li}$ - $n_{Zn}\mu_{Zn}$ - $n_P\mu_P$, where ET is the total energy of the supercell, $n_x$ is the number of x atom in the supercell, and $\mu_x$ is the atomic chemical potential. It has $\mu_{Li} + \mu_{Zn} + \mu_P = \mu_{LiZnP(bulk)}$. Tab.1 shows formation energy for two extreme conditions, i.e., the Li-rich & Zn-rich limit ($\mu_{Li} = \mu_{Li(bulk)}$, $\mu_{Zn} = \mu_{Zn(bulk)}$) and the Li-rich & P-rich limit ($\mu_{Li} = \mu_{Li(bulk)}$, $\mu_P = \mu_{P(bulk,black)}$). The experimental condition can be between these two extreme conditions.

| LiZnP with excess Li | Formation energy (Li-rich & Zn-rich limit) | Formation energy (Li-rich & P-rich limit) |
|---|---|---|
| Interstitial Li (supercell $Li_{28}Zn_{27}P_{27}$) | 2.67 eV | 2.67 eV |
| Li at Zn site & Zn is removed (supercell $Li_{28}Zn_{26}P_{27}$) | 0.45 eV | -0.93 eV |

# Figure Captions

**FIG. 1.** (a) Crystal structure of LiZnP showing [ZnP$_4$] tetrahedral coordination; (b) Lattice constants of Li$_{1+y}$ZnP, Li$_{1.02}$Zn$_{1-x}$Mn$_x$P, and Li$_{1.04}$Zn$_{1-x}$Mn$_x$P for various Li and Mn concentrations. The systematic change of lattice parameters indicates the successful substation of the elements.

**FIG. 2.** DC-magnetization measurements of Li (Zn,Mn)P systems. (a) *M(T)* curves in ZFC and FC modes with *H* = 100Oe for Li$_{1.04}$(Zn$_{1-x}$Mn$_x$)P specimens (no visible difference between ZFC and FC procedures for small coercive fields). Inset shows the temperature dependence of the inverse susceptibility. (b) Top: $T_C$ and $\theta$ for Li$_{1+y}$Zn$_{1-x}$Mn$_x$P; Bottom: $M_{eff}$ and $M_{sat}$ for Li$_{1.04}$Zn$_{1-x}$Mn$_x$P. Dashed lines indicate extrapolation. (c) *M(H)* and *ρ(H)* in Li$_{1.04}$Zn$_{0.9}$Mn$_{0.1}$P show hysteresis, demonstrating ferromagnetism with a small coercive field of about 50Oe.

**FIG. 3.** (a) *ρ(T)* and the inverse carrier concentration for Li$_{1.04}$Zn$_{0.9}$Mn$_{0.1}$P. Inset shows Hall resistivity of Li$_{1.04}$Zn$_{0.9}$Mn$_{0.1}$P, exhibiting p-type carriers with $n_p$=6.1×10$^{16}$cm$^{-3}$, 1.82×10$^{17}$cm$^{-3}$ and 3.27×10$^{17}$cm$^{-3}$ at 100K, 200K and 300K, respectively. (b) Correlation between $T_C$ and the hole concentration for four types of tetrahedrally coordinated DMS systems. It exhibits large difference between Li(Zn,Mn)P and other tetrahedrally coordinated systems. The blue stars represent (In,Mn)As films with electric-field controlled hole concentration. (c) Magnetoresistivity $\rho_H(T)$ of Li$_{1.04}$Zn$_{0.9}$Mn$_{0.1}$P under various field strengths. Inset shows negative magnetoresistance at low temperature.

**FIG. 1.**

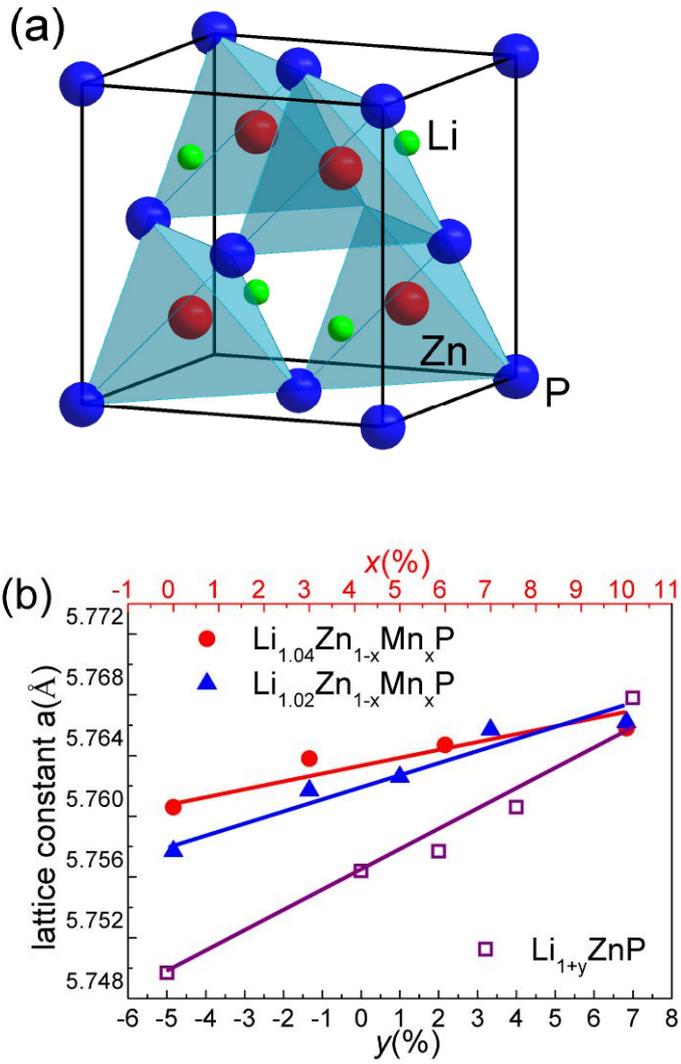

**FIG. 2.**

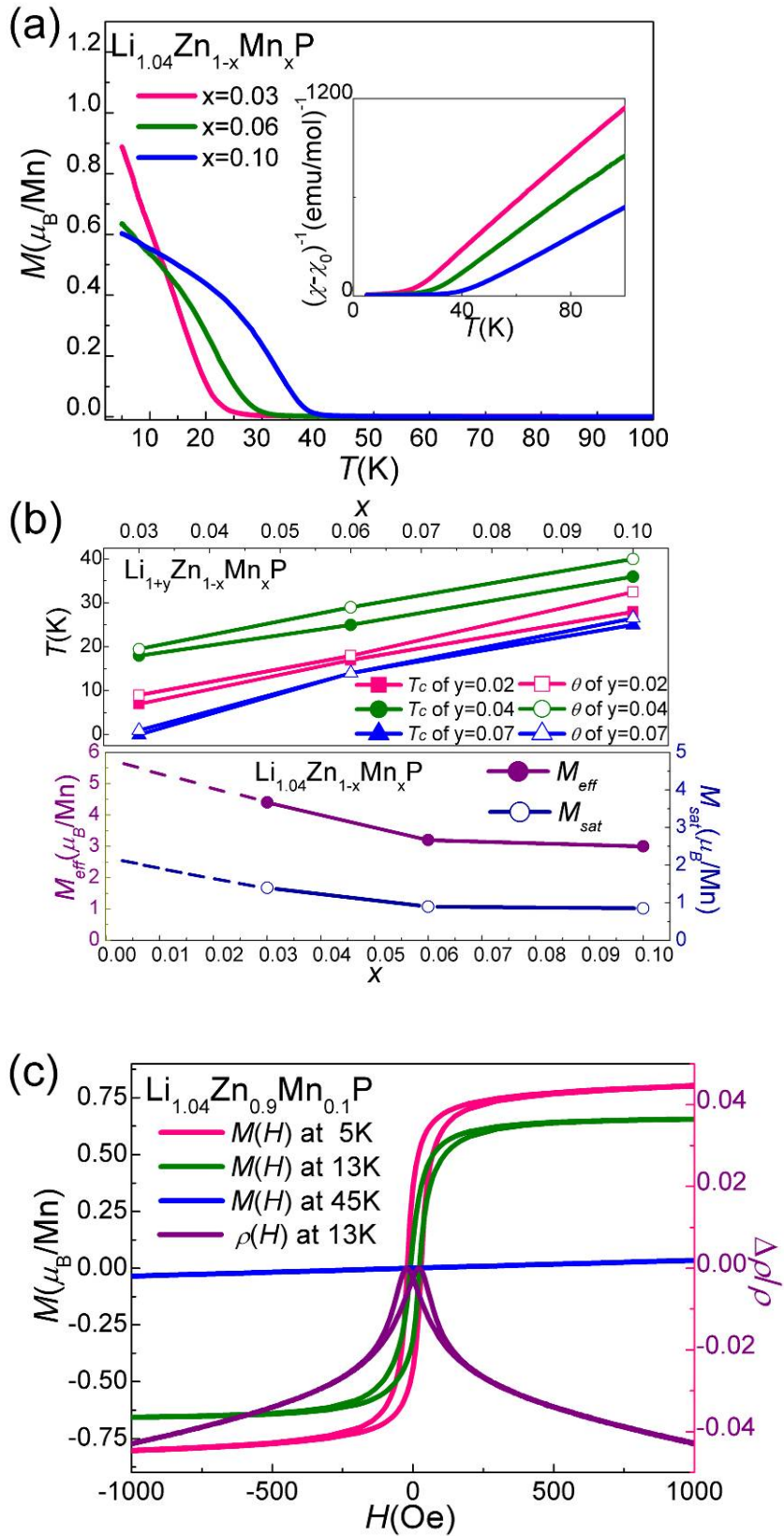

**FIG. 3.**

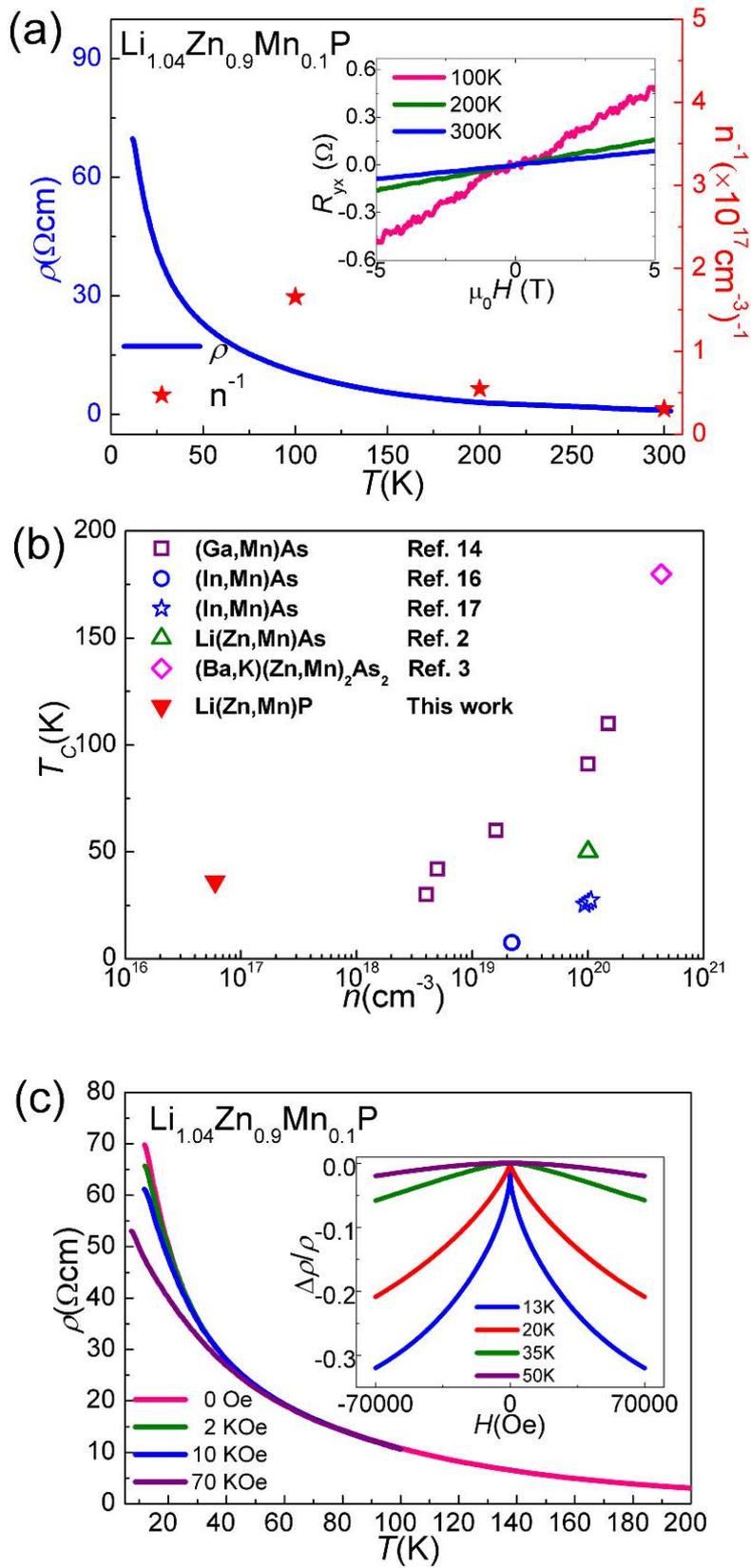